\documentclass[preprint, raferee]{aastex}
\usepackage{epsfig,amsmath,amsfonts,amssymb,graphicx,subfigure,enumitem,color}
\shorttitle{Wave Activity in EK Dra}
\shortauthors{Srivastava et al.}

\begin{document}
\title{First Evidence of the Frequency Filtering of Magnetoacoustic Waves in the flaring star EK Dra}
\author{A. K.~Srivastava}
\affil{Department of Physics, Indian Institute of Technology (BHU), Varanasi-221005, India.}
\author{J. C.~Pandey and Subhajeet Karmakar}
\affil{Aryabhatta Research Institute of Observational Sciences (ARIES), Nainital - 263002, India}
\author{Partha Chowdhury and Y.-J.~Moon}
\affil{School of Space Research, Kyung Hee University, Yongin, Gyeonggi-Do, 446-701, Korea 0000-0001-6216-6944}
\author{Marcel Goossens}
\affil{Centre for mathematical Plasma Astrophysics, Mathematics Department, KU Leuven, Celestijnenlaan 200B bus 2400, B-3001 Leuven, Belgium}
\author{P.~Jel\'inek}
\affil{University of South Bohemia, Faculty of Science, Institute of Physics and Biophysics, Brani\v sovsk\'a 1760, CZ -- 370 05 \v{C}esk\'e Bud\v{e}jovice, Czech Republic}
\author{M.~Mathioudakis}
\affil{Astrophysics Research Centre, School of Mathematics and Physics, Queen's University Belfast, BT7 1NN, UK}
\author{J.G.~Doyle}
\affil{Armagh Observatory, College Hill, Armagh 9DG 73H, N. Ireland}
\author{B.N.~Dwivedi}
\affil{Department of Physics, Indian Institute of Technology (BHU), Varanasi-221005, India.}
\bigskip
\begin{abstract}
Using the data obtained from  XMM-Newton, we show 
the gradual evolution of two periodicities of $\approx$4500 s and $\approx$2200 s in the decay phase of the flare observed in a solar analog EK Dra. The longer period evolves firstly for first 14 ks, while the shorter period evolves for next 10 ks in the decay phase. We find that these two periodicities are associated with the magnetoacoustic waves triggered in the flaring region. The flaring loop system shows cooling and thus it is subjected to the change in the scale height and the acoustic cut-off period. This serves to filter the longer period magnetoacoustic waves and enables the propagation of the shorter period waves in the later phase of the flare. We provide the first clues of the dynamic behaviour of EK~Dra's corona which affects the propagation of waves 
and causes their filtering.

\end{abstract}

\keywords{ magnetohydrodynamics (MHD)  stars: coronae  stars: flare  stars: individual (EK Dra) stars: oscillations (including pulsations)  waves} 
\bigskip

\section{Introduction}

Magnetohydrodynamic (MHD) waves and oscillation are important physical phenomena arose in the magnetized plasma that provide a  significant diagnostics (e.g., magnetic field, density and magnetic scale heights, transport phenomena, etc) of the localized solar and stellar coronae \citep{2005LRSP....2....3N, 2009SSRv..149..119N}. The waves and oscillations are ubiquitous in the solar atmosphere, and are studied both observationally \citep[e.g.,][references cited therein]{2004A&A...421L..33W,2008MNRAS.388.1899S, 2009Sci...323.1582J, 2011ApJ...736..102A, 2012A&A...545A.129W, 2013ApJ...777...17S,2017NatSR...743147S} and theoretically \citep[e.g.,][references cited therein]{2005A&A...435..215K, 2011SSRv..158..289G, 2013MNRAS.434.2347J, 2014RAA....14..805P, 2015A&A...581A.131J, 2016A&A...590A...4K} in different magnetic structures. There are several pieces of evidence that they also exist in the atmosphere of similar type magnetically active stars \citep[e.g.,][and references cited there]{2003A&A...403.1101M,2005A&A...436.1041M,2006A&A...456..323M,2009ApJ...697L.153P,2013ApJ...778L..28S,2018MNRAS.475.2842D}.
These waves or periodic magnetic reconnection may evolve there during powerful flaring energy release, and can modulate various emissions across the electromagnetic spectrum of star \citep[e.g.,][and references cited therein]{2005A&A...432..671S,2008MNRAS.387.1627P,2015ApJ...813L...5P,2016ApJ...830..110C}.

Likewise solar atmosphere, these MHD waves may most likely excite in the structured magnetic tubes of the stellar atmosphere. Solar atmosphere offers significant coupling, conversion, reflection of MHD modes due to its complex magnetic field and plasma structuring, therefore, imposes a constraint on propagation of these waves and transport of associated power (e.g., McAteer et al. 2003; Schrijver \& Title, 2003; De Pontieu et al., 2004; Centeno et al., 2006; Vecchio et al., 2007; Khomenko \& Cally, 2012, and references cited therein).  

In the solar atmosphere, most powerful acoustic oscillations are associated with five min photospheric motions, which may convert into magnetoacoustic waves into the region where plasma beta becomes one (Fedun et al., 2009). However, the propagation of 5 min acoustic oscillations are highly debated, and depends upon certain properties of the plasma as well as magnetic field of the localized solar atmosphere (e.g., De Pontieu et al., 2004; Khomenko et al., 2008; Yuan et al., 2014). The structured solar atmosphere works as a wave-guide for various MHD modes, and also serve as a frequency filter transporting selective frequencies and associated energies (e.g., Fedun et al., 2011; Murawski et al., 2016). However, this fact has never been explored as far as the stellar coronae of similar stars are considered. 

In this paper, we have investigated the quasi-periodic oscillation detected during the decay phase of a flaring event in an active star EK Dra.  EK Dra is a young solar analog which provides an opportunity to study the magnetic activity of the infant Sun. Except the age, EK Dra has similar properties to that of the Sun. It is a G1.5V star with 2.75 d rotation period  and located at 34 pc away from the earth. We study for the first time that the flaring atmosphere in EK Dra filters the evolved frequencies, and thus has the structured atmosphere likewise Sun. The observational data and reduction are described in section 2. Results are outlined in section 3 and Section 4 depicts the theoretical interpretations. Discussion along with the conclusions are given in the last section.

\begin{figure}
\centering
\includegraphics[width=90mm, angle=-90]{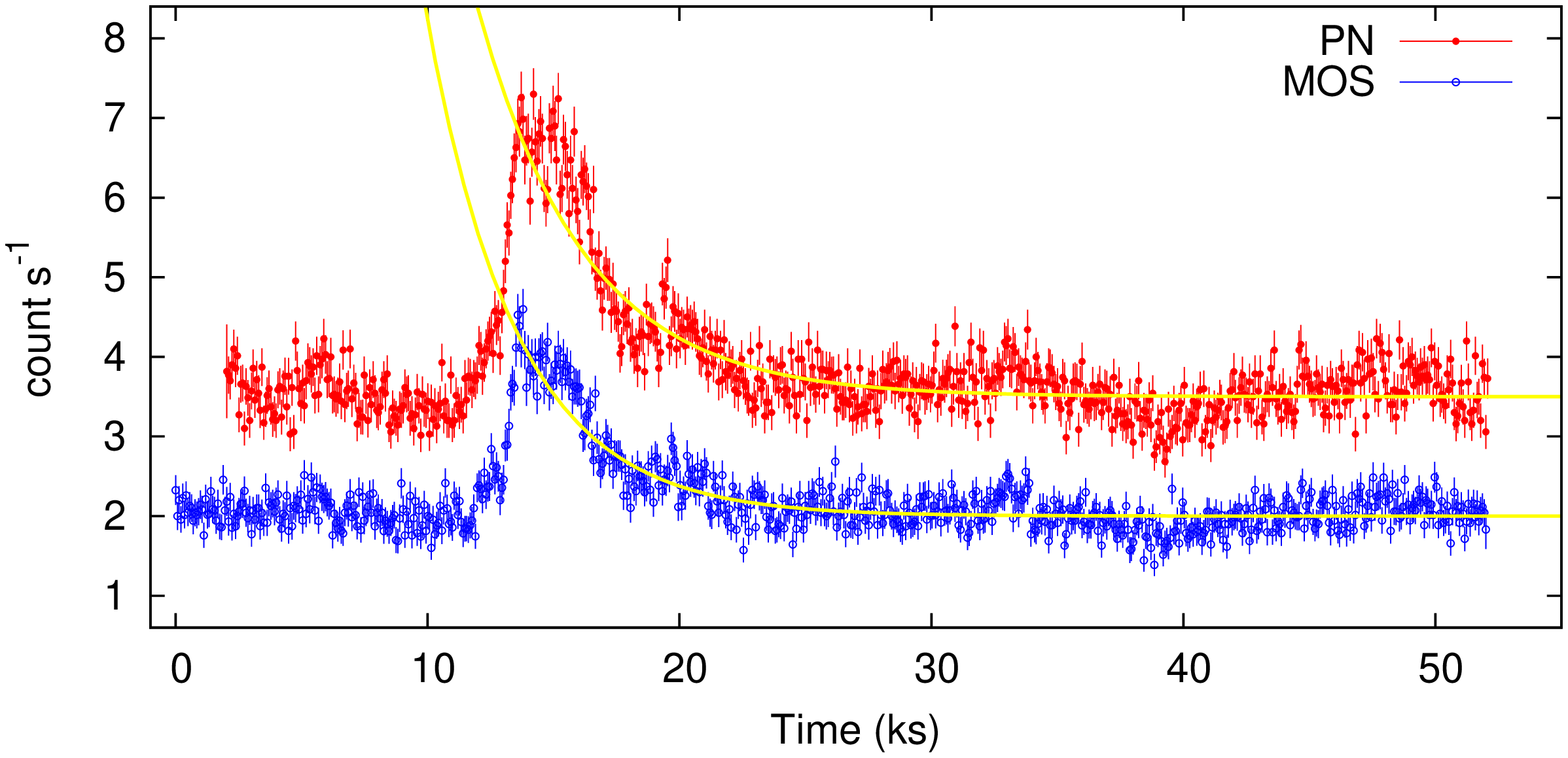}
\includegraphics[width=90mm, angle=-90]{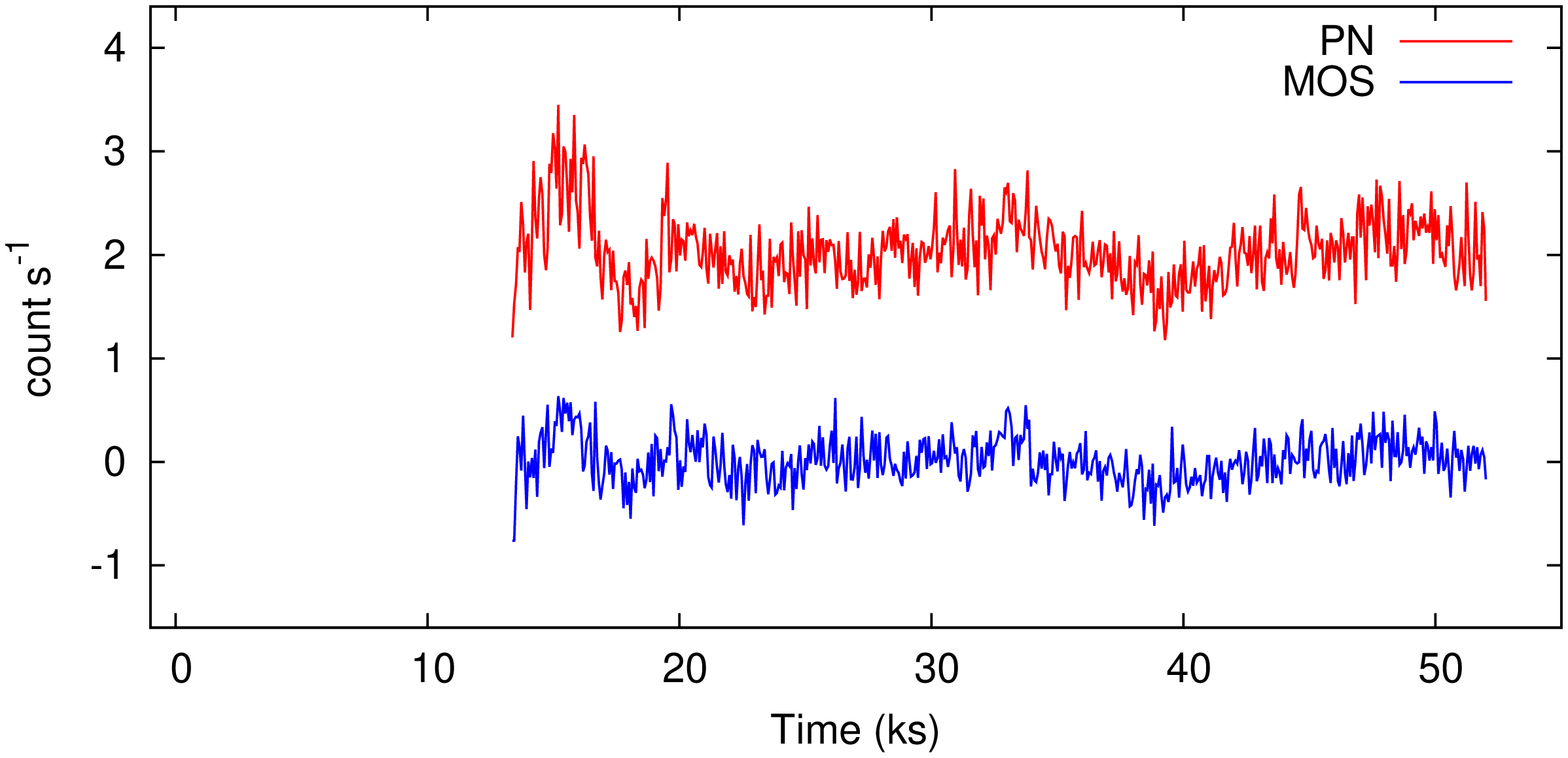}
\caption{Top panel: Background subtracted X-ray light curves of EK Dra in the 0.3-10.0 keV energy band,  upper curve (red color) is for PN data while lower curve (blue color) is for MOS data with time bins of 70s. The continuous line (yellow color) is the best fit exponential function of the decay phase. Bottom panel: The detrended light curve from peak to the decay phase. The PN count rates are adjusted by adding 2 count s$^{-1}$.}
\label{fig:lc}
\end{figure}

\section{Observations}

 EK Dra had been observed for 54.9 ks by the XMM-Newton satellite using the European Photon Imaging
Camera (EPIC) and Reflection Grating Spectrometer(RGS) at 14:01:58 UT on December 30, 2000.
The EPIC is composed of three CCDs behind three X-ray telescopes (Jansen et al. 2001); the twin metal oxide semiconductor (MOS) CCDs, MOS1 and MOS2 (Turner et al. 2001), and one p-n junction CCD, PN (Str{\"u}der et al. 2001). The RGS is composed of two identical grating spectrometers, RGS1, and RGS2, behind different mirrors (den Herder et al. 2001). The data were reduced using the standard  Science Analysis System (SAS) of XMM-Newton version 16.0.0 with updated calibration files. X-ray light curves were generated from on-source counts obtained from circular regions with a radius of 40 arcsecs around the source. The background
was chosen from source-free regions on the detectors surrounding the source. The detailed spectral analyses of EPIC and RGS data were performed by Scelsi et al. (2005). 

\section{Results}
\subsection{X-ray light curve and loop parameters}

Top panel of Fig.~\ref{fig:lc} shows the background subtracted X-ray light curves of EK Dra in the 0.3-10.0 keV energy band. A sudden enhancement is followed by a gradual decay indicating a flare during the observations. The intensity reached a maximum value at 13.4 ks from the start of the MOS observation. The flare lasted for $\sim 10$ ks. The continuous line shows the best fit exponential function and we found  the e-folding decay time ($\tau_d$) of  $4221 \pm 98~{\mathrm{seconds}}$. The detrended light curve during the decay phase is shown in the bottom panel of the  Fig.~\ref{fig:lc}.

Time-resolved spectral analysis of the observed flare by Scelsi et al. (2005) showed that the temperature and emission measure during the flare follow the behaviour of light curve. The temperature had a peak before the emission measure. The average maximum temperature was found to be $42^{+1.6}_{-1.0}$ MK. The maximum temperature $T_{max}$  was calculated from the observed maximum temperature using method of Reale (2007), and it is estimated to be  $91\pm27$ MK.
 Scelsi et al. (2005) have also  derived the equilibrium temperature of the flaring plasma as 8.4 MK at the end of the decay phase.
According to the state-of-art hydrodynamic model of Reale et al. (1997), the loop length is estimated as

\begin{equation}
 L ~(~cm~)~ = 5.27\times10^3    \frac{\tau_d ~ \sqrt{T_{max}}}{F(\zeta)} ~~~~~~ {\rm     for  ~0.35 <\zeta < 1.6}
\label{eq:loop}
\end{equation}

\begin{figure*}
\includegraphics[width=175mm]{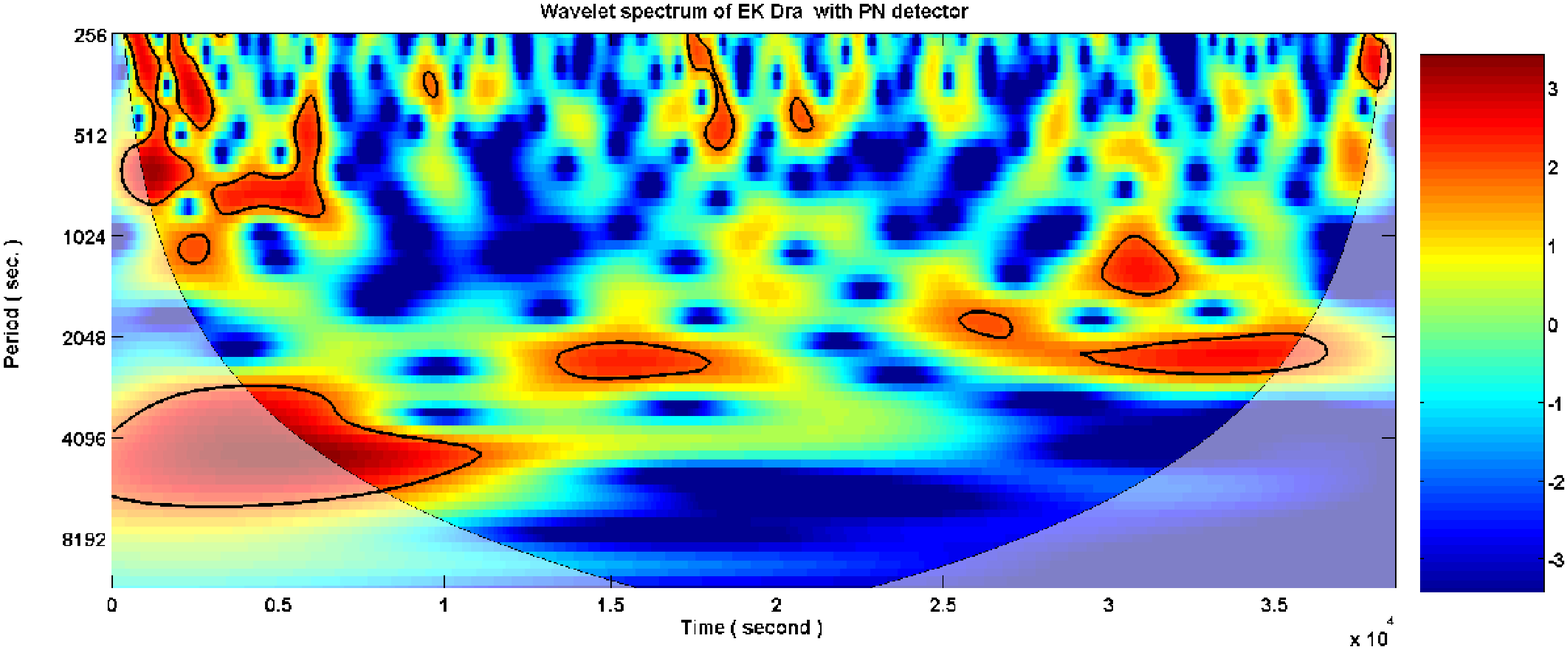}
\includegraphics[width=175mm]{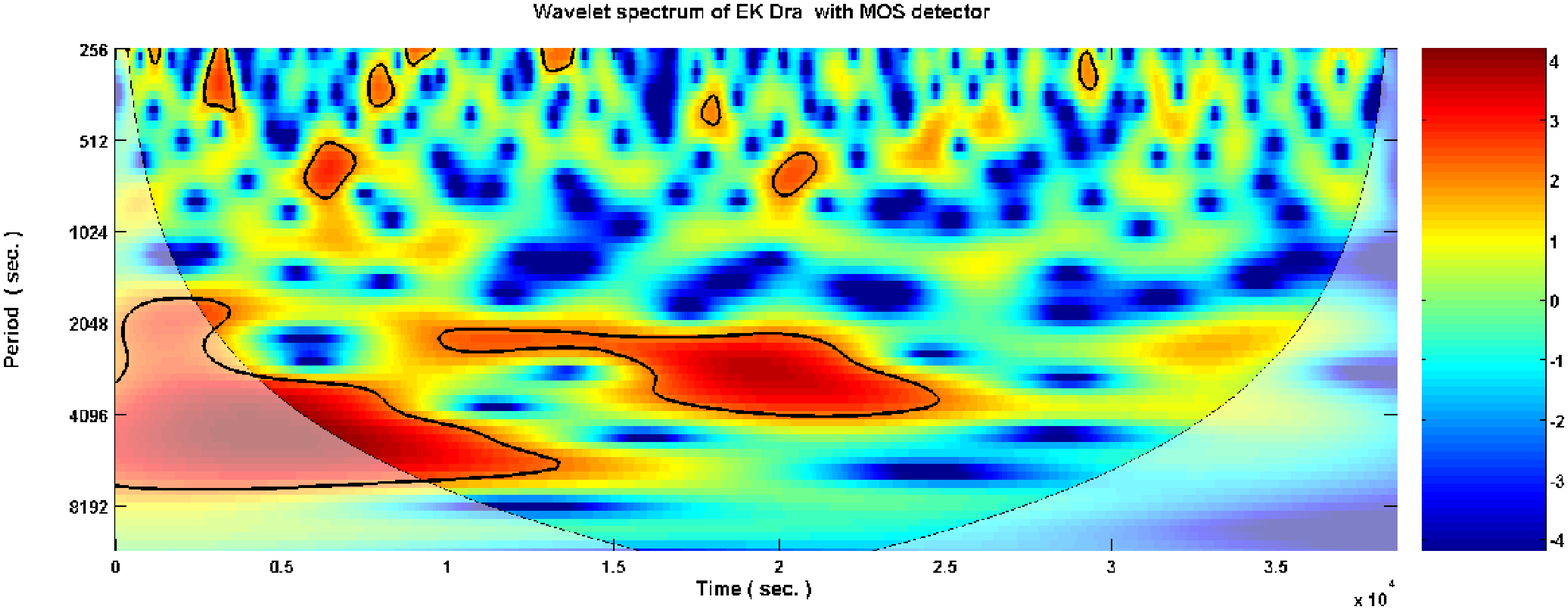}
\caption{Wavelet power spectrum corresponding to the detrended X-ray light curves (Fig.~1) of EK Dra in the 0.3-10.0 keV energy band.}
\label{fig:wavelet}
\end{figure*}

The dimensionless correction factor is $F(\zeta) = \frac{0.51}{\zeta-0.35}+1.36$; where $\zeta$ is the slope of the log $\sqrt{\rm EM}    $ \textendash ~log $T$ diagram (Reale 2007). Scelsi et al. (2005) have derived a value of 1.0 for $\zeta$ using two data points during the decay of the flare. As temperature and emission measure are decaying during the post flare phase, therefore, by considering all four data point during this phase (see Table B.1 and Figure B.1 of of Scelsi 2005), we derived the value of $\zeta$ as $1.0\pm 0.5$. Using equation \ref{eq:loop}, we derive a loop length of $9.9 \pm 3.1 \times 10^{10}$ cm. Table \ref{tab:parameter} summarizes the derived loop parameters.

The electron density of EK Dra  was derived by using He-like triplets from O{\sc vii}. The most intense He-like lines correspond to the transitions between the $n = 2$ shell and the $n = 1$ ground state shell. The excited state transitions  to the ground state are called resonance ($r$), inter-combination ($i$) and forbidden ($f$) lines. In the X-ray spectra, the ratio of fluxes in forbidden  and inter-combination lines ($R = f/i$) is  sensitive to the electron  density (Porquetd et al. 2001). We have used the CHIANTI atomic  database version 8.0.2 (Dere et al. 1997; Delzanna et al. 2015) to derive  the density from  $R-$ratios. Using $R = 3.0\pm1.7$ and a temperature of 1.5 MK (see Table 5 of Scelsi et al. 2005 also), the density was estimated to be $2.5\times10^{10}$ cm$^{-3}$. 
For the fully ionized plasma the pressure p ($=2n_ekT_{max}$) was derived to be 628 dyne cm$^{-2}$. Using the relation, $p=B^2/8\pi$, the minimum magnetic field required to confine the plasma was derived to be 126 G.  


\begin{table}
\caption{Loop parameters of EK Dra}\label{tab:parameter}
\begin{tabular}{ll}
\hline
Parameters & Value\\
\hline
Decay Time ($\tau_d$)  & $4234\pm98$s \\
$T_{max}$              &$9.1\pm2.7 \times 10^7$ K\\
Loop Length(L)          &$9.9\pm3.1 \times 10^{10}$ cm\\
Electron Density ($n_e$) & $2.5\times10^{10}$ cm$^{-3}$\\
Pressure ($p = 2n_ekT_{max}$)             & 628 dyne cm$^{-2}$\\
Magnetic Field (B)$^1$   & 126 Gauss\\
\hline
\end{tabular}
~\\
$^1$ minimum magnetic field to confine the plasma 
\end{table}

\begin{figure*}
\includegraphics[width=160mm]{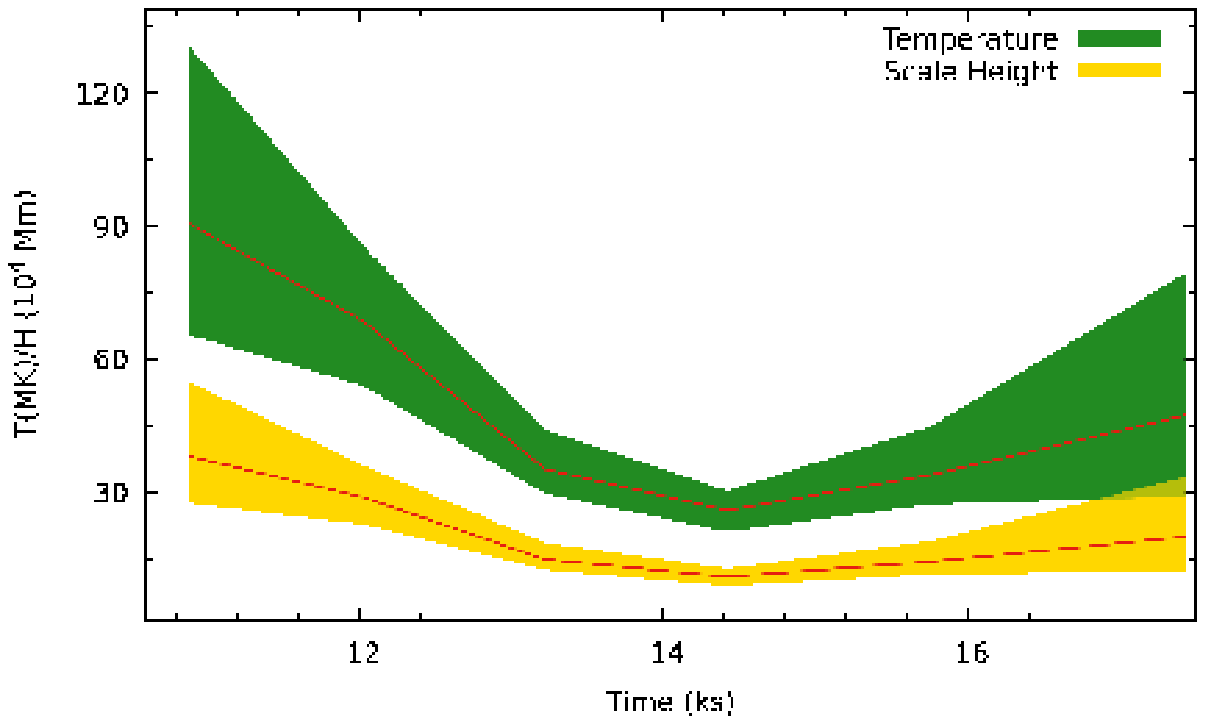}
\caption{The temperature and scale height variation in the decay phase of the flare.}
\label{fig:wavelet}
\end{figure*}

\subsection{Power spectra and associated results} 

We use the wavelet analysis code (Torrence \& Compo 1998) to examine the statistically significant periodicities in the EK Dra time-series. 
Using wavelet analysis, the search for periodicities in XMM-Newton X-ray light curves is carried out by a time-localized function, which is continuous in both frequency and time. We use the Morlet wavelet function in the present time-series analysis which is defined as follows
\begin{equation}
\psi_{t} (s) = \pi^{-1/4} e^{i \omega t} \ e^{ \left (\frac {-t^{2}}{2s^{2}} \right)}
\end{equation}

\noindent
where $t$, $s$, $\omega$ and $\pi^{-1/4}$, are time, wavelet scale, oscillation frequency, and the normalization constant, respectively. 
The Morlet wavelet is a sinusoidal function modified by a Gaussian envelop. In the Morlet wavelet, the Fourier period ($P$) is related to
the wavelet scale ($s$) by the relation P = 1.03 s. The wavelet is convolved with
the chosen observed time series to further determine the contribution of
frequency within this time series, which matches the
sinusoidal portion by varying the scale of the wavelet
function. This method provides the power spectrum of
the oscillations in different X-ray light curves of the flaring 
epoch in EK Dra as observed by XMM-Newton. The Morlet wavelet consists of an $"$edge effect$"$
, which is present in the analyses of time series data. However,
this effect is significant only in regions lying
within the cone-of-influence (COI) which demarks where 
possible estimated periods, very close to either the measurement
interval or the maximum length of the time series,
cannot be convincingly detected.

Fig.~2 shows the wavelet spectrum of the 70 sec bin detrended X-ray light curve of the flaring epoch on EK Dra (see Fig.~1) in its decay phase with PN and MOS detectors. The light curves are detrended by fitting an exponential function to remove the long term trend of flare variations (Fig.~1, bottom panel). The intensity wavelets show the evolution of two periodicities at $\approx$4500 s and $\approx$2200 s respectively. It should be also noted that these two periodicities are detected in the X-ray light curves observed by two detectors (PN \& MOS) simultaneously  onboard XMM-Newton, therefore, they must be considered as reliable periods present in the decay phase of the flare. Moreover, a unique property of these two periodicities is that they are not evolving simultaneously. 
Instead, the long period of 4500 s evolves firstly up to a duration of $\approx$14 ks, and thereafter, the lower period of 2200 s evolves. The power associated with both periods is globally distributed as they repeat $>$3 cycles in the total time epoch of the observations (Fig.~2).
Moreover, the evolved periodicities are consistent in the X-ray light curves observed at different bins in both the detectors PN \& MOS.  

\section{Theoretical Interpretation}

The quasi-periodicities of $\approx$4500 s and $\approx$2200 s evolve in the decay phase of the stellar flare, and the second 
periodicity of 2200 s is almost half of the first period 4500 s.
The most likely explanation for these periodicities is that they are associated with the evolution of MHD waves.
An investigation of the wavelets diagrams (Fig.~2) shows that both periods evolve gradually. The 4500 s period  evolves firstly for 14 ks during the decay phase of the flare, while 2200 s period evolves thereafter for 10 ks. Both periods are distributed for $>$ 3 cycles in the intensity 
wavelet exhibiting their global nature and associated power distribution. Therefore, these periods are not associated 
with any transient reconnection process which may lead to the localized burst and enhancement of the intensity in real time
series as well as power in the localized Fourier domain (e.g., Doyle et al., 2018).

\begin{figure*}
\includegraphics[width=85mm, angle=-90]{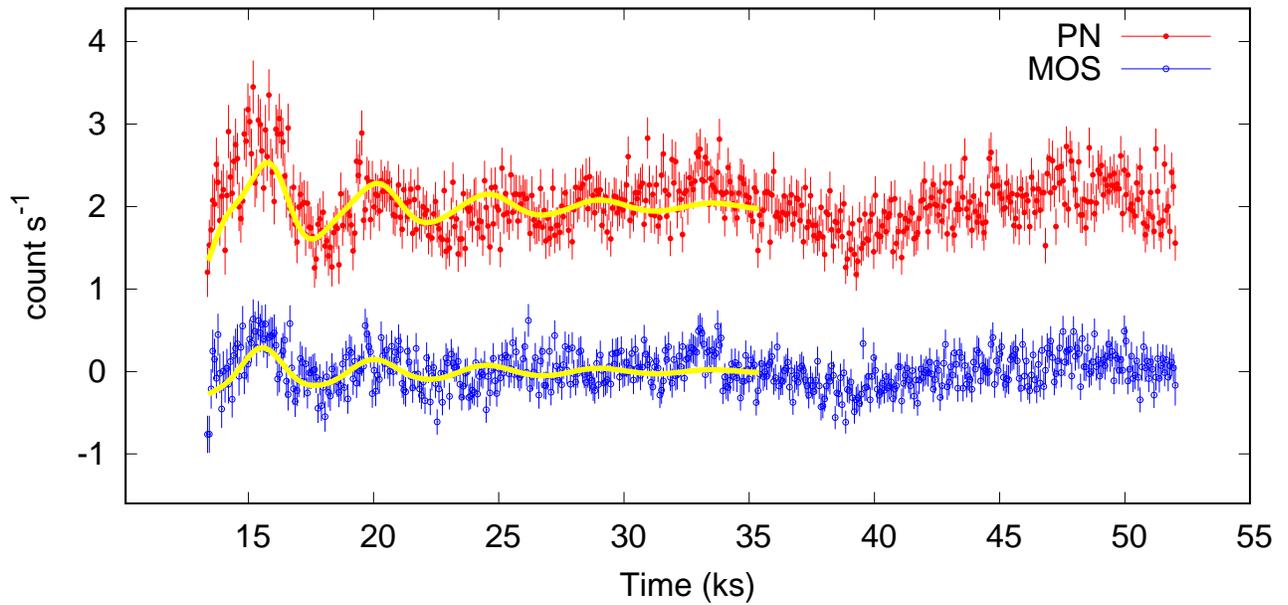}
\caption{The fitted light curve (yellow) with a combination of 4500 s and 2200 s periods over the 
emission in the decay phase of the flare.}
\label{fig:wavelet}
\end{figure*}

We suggest that the 4500 s period may be associated with the evolution of magnetoacoustic waves in the flaring loop
system of EK Dra. Interestingly, the longer period switched towards a comparatively shorter 
period in due course of time in the observational base-line, and they do not evolve simultaneously. This is 
the first most likely evidence that the magnetized atmosphere  of a solar analogue acts like a frequency filter and it is 
well structured. We have analyzed how the flaring atmosphere in the atmosphere of EK Dra is filtering the wave frequencies.

Fig.~3 shows the variation of the temperature of the atmosphere in the decay phase of the flare. Assuming 
hydrostatic equilibrium, the scale height (H) of the atmosphere depends on temperature (T) as H=k$_{B}$T/$\mu$g,
and changes as the temperature changes. Here, k$_{B}$ is the Boltzman constant, $\mu$ is the mean molecular weight in the terms of proton's rest mass and g the surface gravity of a star. For EK Dra, we took  g = 329 ms$^{-2}$ (Scelsi et al. 2005) and $\mu$ = 0.62 (for solar abundances).  It should be noted that in any particular instance, the flaring loop system
is considered to be an isothermal loop system, although it is a cooling as far as the elapse of time epoch is concerned.
It is shown in Fig.~3 that both the scale height and temperature drop to their
minimum value up to $\sim$14 ks after the flare peak, and remain almost at their minimum values during the course of the observations
in the decay phase of the flare. This is the time when frequency filtering is evident in the wavelet diagrams (Fig.~2).  Here, the average temperature was converted to maximum temperature as per the relation $T = 0.13 T_{obs}^{1.16}$ (see Reale 2007).
The acoustic cutoff frequency of the 
atmosphere depends upon the sound speed and scale height as $\Omega_{ac}$=c$_{s}$/2H. Therefore, the ratio of the acoustic cut-off
frequency near the flare peak ($\Omega_{ac1}$), and after 14 ks ($\Omega_{ac2}$) in the decay phase of the flare will be 

\begin{equation}
\frac{{\Omega_{ac1}}}{{\Omega_{ac2}}}=\sqrt{\frac{T1}{T2}}\times\frac{H2}{H1}\,,
\end{equation}

which becomes 0.6.  For a  maximum temperature of 91 MK at the flare peak, we have estimated the acoustic cut-off period of the flaring atmosphere as 5268 s. As the time elapsed, the loop system is cooled
down and after 14 ks the acoustic cut-off period will be reduced to the value of 2820 s. 

This clearly demonstrates that the broad-band (multiple) frequency waves are excited in the flaring loop system during the onset 
of the flare. In the beginning, the most dominant period evolves at the longer period, and the 4500 s periodicity is detected 
in the intensity wavelet. The higher acoustic cut-off period of 5268 s easily allows the lower periods having dominant power 
to propagate in the flaring loop system that is evident in the wavelet power spectrum (Fig.~2).
As the flaring atmosphere cools down, plasma structuring changes cause the reduced cut-off period
up to a value of 2820 s. Therefore, it eliminates the dominant longer period waves having periods more than the acoustic cut-off period,
however, it allows the lower periods (here 2200 s) to propagate in the atmosphere. Once the dominant longer periods are filtered by
the atmosphere, the lower periods having the significant power becomes evident in the intensity wavelet (Fig.~2). 

Fig.~4 shows the fitted oscillatory (yellow) curves on the actual emissions in the decay phase of the flare.
The fitted curve is the combination of two dominant signals with periods of 4500 s and 2200 s respectively. This curve is based on the variation of exponentially decaying harmonic function with the detrended light curve. The function is described as follows

\begin{equation}
I(t)=\sum_{n=1}^{2}{A_{n}Cos}\Bigg(\frac{2\pi}{P_{n}}.t+\phi_{n}\Bigg)e^{\frac{-\delta}{P_{n}}.t}\,,
\end{equation}

where A$_{n}$ and $\phi_{n}$ are amplitudes and phases corresponding to the oscillatory periods, 
P$_{n}$ (4500s and 2200 s), respectively, and $\delta$ is the damping factor. The best fit of the function
(cf., yellow curve in Fig.~4) gives $\delta=0.64\pm0.013$, which indicates a damping. This curve fits well the
observed light curve by XMM-Newton detectors consisting of the signals corressponding to the detected periodicities (cf., Figs. 1-2). The synthesized curve shows the decay, which is similar to
the observed decay when the initial amplitude of the oscillations reduces to its 1/e value in the observational base-line. It should be noted that the synthesized signal with the 
combination of two observed periodicities fits well the observed light curves up to 25 ks of the 
observations, and produces a consistent result.
This indicates that multiple periodicities are excited in the flaring loop system episodically
during the flare energy release, and these are related with the magnetoacoustic wave modes. However, the structured  flaring atmosphere works as a natural filter
for the excited waves. Initially the dominant long period waves (P$_{1}$=4500 s) are detected in the observational baseline,
and thereafter, they get attenuated by the medium and only short period waves (P$_{1}$=2200 s) are allowed.

There may also be another possibility that these detected periods are associated with the multiple harmonics of the standing slow 
magneto-acoustic waves. The 4500 s periodicity may arise due to the fundamental mode of slow magnetoacoustic 
modes which can perturb the density and can also modulate the emissions of the flaring loops where they excite.
While, the 2200 s periodicity may arise due to the first overtone evolved in the same flaring loop system.
The loop length is estimated as 1$\times$10$^{11}$cm (cf., Table~1), therefore, the phase speed of the fundamental 
mode of the oscillations is V$_{ph}$=2L/P$\approx$440 km s$^{-1}$ that is indeed a slow mode while 
we compare its phase speed with the local sound speed 
at average flare temperature in its decay phase (cf., Fig.~3).
However, these periods do not evolve simultaneously as a multiple wave harmonics. Instead, they evolve
one by one (from longer to shorter period) in the decay phase of the flare when flaring loops are being cooled (Fig.~3). The most likely
scenario is that due to the injection of flare energy, the pressure (and thus density) of the hot loops are perturbed causing  
the evolution of various slow magnetoacoustic waves. Since the loop's environment is continuously changing in the decay phase of the flare,
therefore, the associated acoustic cut-off frequency also changes, which allows  certain frequencies to propagate in most predominant ways.

\section{Discussion and Conclusions}

In the solar coronal loops, due to the injection of flare energy, the evolution of slow magnetoacoustic waves are 
one of the most common physical scenario (e.g., Kumar et al., 2016; Nakariakov \& Zimovets, 2011; Fang et al., 2015, and references
cited therein). Apart from propagating waves, the standing slow waves are also well observed in cool and hot flaring loops in 
the solar atmosphere (e.g., Ofman \& Wang, 2002; Wang \& Solanki, 2004; Srivastava et al., 2010; Kumar et al., 2015, and 
references cited therein). These waves are also observed in the stellar coronae (Srivastava et al., 2013). 

In the present paper,
we firstly observe the existence of slow magnetoacoustic waves over a range of periods in the flaring region. The highly dynamic
atmosphere of EK Dra enables selective propagation of these waves, while it attenuates other frequencies. Similar behaviour is well 
observed in the Sun's atmosphere where complex structuring of its atmosphere causes filtering of the magnetoacoustic waves (Yuan et al., 2014;
Murawski et al., 2016). The frequency filtering imposed above the localized flaring corona of EK Dra may indicate the behaviour of the high pass filter, and in fact such filter may not amplify incident powers. Therefore, both the periods were simultaneously present in first 14 ks of the flaring epoch in its decay phase which is seen in MOS signals (Fig.~1, bottom-panel), however, 4500 s period and related power were dominant in that phase. 
As the temperature minimum condition emerges in the flaring region due to the gradual cooling, the changing background atmosphere filters the longer period waves and allow the passage of comparatively smaller periods selectively. We provide the first evidence of such physical behaviour in the magnetized corona of the sun-like star EK Dra. Our present study provides a new insight that the EK Dra solar analog may have very similar atmosphere and wave activity as what we observe on the Sun. However, more investigations need to be carried out about this star using multiwavelength observations to understand 
magnetohydrodynamic (MHD) wave activity in its corona. This will also provide the clues that how its atmosphere is channeling the wave 
energy, and affecting the wave propagation properties.
 
\section{Acknowledgment}

This work uses data obtained by XMM-Newton, an ESA science mission with instruments and contributions directly funded by ESA Member States and the USA (NASA). 
AKS and MM acknowledge the support of UKIERI Project to support their joint research.
PJ acknowledges support from Grant 16-13277S of the Grant Agency of the Czech Republic. PC and YJM acknowledge the support by the BK21 plus program from the National Research Foundation of Korea to Kyung Hee University, Korea. JCP, AKS, and SK also acknowledge the DST-RFBR grant INT/RUS/RFBR/P-271.

\end{document}